\documentclass[twocolumn]{aastex631}
\usepackage{amsmath}
\newcommand{\m}{$M_{\rm{ZAMS}}$}

\begin{document}

\title{Progenitor constraint with circumstellar material for the magnetar-hosting supernova remnant RCW~103 }

\author{Takuto Narita}
\affiliation{Department of Physics, Kyoto University Kitashirakawa Oiwake-cho, Sakyo, Kyoto, 606-8502, Japan}

\author{Hiroyuki Uchida}
\affiliation{Department of Physics, Kyoto University Kitashirakawa Oiwake-cho, Sakyo, Kyoto, 606-8502, Japan}

\author{Takashi Yoshida}
\affiliation{Yukawa Institute for Theoretical Physics, Kyoto University, Kitashirakawa Oiwake-cho, Sakyo, Kyoto 606-8502, Japan}

\author{Takaaki Tanaka}
\affiliation{Department of Physics, Konan University, 8-9-1 Okamoto, Higashinada, Kobe, Hyogo 658-8501, Japan}

\author{Takeshi Go Tsuru}
\affiliation{Department of Physics, Kyoto University Kitashirakawa Oiwake-cho, Sakyo, Kyoto, 606-8502, Japan}

\begin{abstract}
Stellar winds blown out from massive stars ($\gtrsim 10M_{\odot}$)  contain precious information on the progenitor itself, and in this context, the most important elements are carbon (C), nitrogen (N), and oxygen (O), which are produced by the CNO cycle in the H-burning layer.
Although their X-ray fluorescence lines are expected to be detected in swept-up  shock-heated circumstellar materials (CSMs) in supernova remnants (SNRs), particularly those of C and N have been difficult to detect so far.
Here, we present a high-resolution spectroscopy of a young magnetar-hosting SNR RCW~103 with the Reflection Grating Spectrometer (RGS) onboard XMM-Newton and report on the detection  of \ion{N}{7} Ly$\alpha$ (0.50~keV) line for the first time.
By comparing the obtained abundance ratio of N to O (N/O$=3.8 \pm{0.1}$) with various stellar evolution models, we show that the progenitor of RCW~103 is likely to have a low-mass (10--12~$M_{\odot}$) and medium-rotation velocities ($\lesssim 100~\rm{km~s^{-1}}$).
The results also rule out the possibility of dynamo effects in massive ($\geq35~M_{\odot}$) stars as a formation mechanism of the associated magnetar 1E~161348$-$5055.
Our method is useful for estimating various progenitor parameters for future missions with microcalorimeters  such as XRISM and Athena.
\end{abstract}


\section{Introduction} \label{sec:intro}
 Circumstellar material (CSM) around core-collapse (CC) supernova remnants (SNRs) provides us with important clue to understanding their progenitors \citep[e.g,][]{Dwarkadas_2005, Dwarkadas_2007, Sapienza_2022}, particularly of massive stars \citep[$\gtrsim10M_{\odot}$;][]{Smartt_2009}.
This is because elemental abundances in the CSM directly reflect physical conditions of progenitors: e.g., zero-age main-sequence mass (\m), initial rotation velocity, and effects of convective overshoot \citep{Maeder_2014}.
In this context, lighter elements, such as nitrogen (N), oxygen (O), and carbon (C), are the most critical since they are produced by the CNO cycle in the H-burning layer.
Among them, N is produced by consuming C and O \citep[e.g.,][]{Maeder_1983}, and then is mixed to the star's surface by the convection and centrifugal force \citep[e.g.,][]{Przybilla_2010}.
Since N-rich materials are  blown out by  radiation pressure  at a stage of red super giants (RSGs) and/or Wolf-Rayet stars (WRs) \citep[e.g.,][]{Owocki_2004},  abundance ratios of N to O (N/O) and N to C (N/C) in the CSM are ideal probes to constrain the progenitors of CC SNRs.
 
While the CNO elements contain fruitful information on progenitors, it is technically difficult to detect their X-ray fluorescence lines in the soft bands (below $\sim0.5$~keV; especially C and N) due to the lack of sensitivity and energy resolution of current detectors.
For estimating \m~and explosion energies of SNRs, many previous studies have alternatively focused on abundances of magnesium (Mg), silicon (Si), sulfur (S), and iron (Fe) in ejecta \citep[cf.][]{Sukhbold_2016, Katsuda_2018} since their fluorescence lines are commonly dominant in SNRs.
Several observations of Galactic SNRs, however, enable detection of \ion{C}{6} Ly$\alpha$ (0.36~keV) and \ion{N}{7} Ly$\alpha$ (0.50~keV) by using the Reflection Grating Spectrometer (RGS)  onboard XMM-Newton \citep[e.g.,][]{Uchida_2019, Kasuga_2021}.
They targeted clumpy structures to avoid degrading energy resolution of the grating spectrometer.
This method will become useful when we attempt to measure abundances of  the CNO elements contained in CSM.

Here we focus on a young Galactic CC SNR, RCW~103, which has a clumpy X-ray morphology, hosting a magnetar candidate 1E~161348$-$5055.
The age of the remnant is estimated to be around 2000~yr by comparing an optical image with a photographic plate \citep{Carter_1997} and the distance is estimated to be 3.1~kpc from a systemic velocity of the \ion{H}{1} line \citep{Reynoso_2004}.
While the origin of 1E~161348$-$5055, which shows an extremely long periodicity \citep[6.67~h;][]{DeLuca_2006}, is currently unknown, a progenitor  \m \ of RCW~103 is also controversial: 18$M_{\odot}$ \citep{Frank_2015}, 12--13$M_{\odot}$ \citep{Braun_2019}, or $<13M_{\odot}$ \citep{Zhou_2019}, which are derived from abundance patterns of the ejecta.

In this paper,  we investigate a CSM-dominant outer region of RCW~103 (\S~\ref{sec:obs} and \ref{sec:ana_res}) and estimate physical conditions of progenitors, such as \m \  and the initial rotation velocity (\S~\ref{sec:evi_com} and \ref{sec:pro}).
We also discuss the relation between the progenitor star and a formation process of the magnetar 1E~161348$-$5055 in \S~\ref{sec:mag} and summarize our conclusions in \S~\ref{sec:con}.
Throughout the paper, the age and distance of RCW~103 are assumed to be 2000~yr \citep{Carter_1997} and 3.1~kpc \citep{Reynoso_2004}.
Errors of parameters are defined as 1$\sigma$ confidence intervals.

\section{Observations}\label{sec:obs}
RCW~103 was observed with XMM-Newton four times as listed in Table~\ref{obs}.
For the following analysis, we only used the RGS and the European Photon Imaging Camera (EPIC) data of these observations. 
A nearby blank-sky observation (see Table~\ref{obs}) was applied for estimating the background emission. 
Since the pn-CCD camera was operated in the Small Window mode throughout the observation and did not cover the shell regions of the SNR, we did not use the pn data in our analysis.
The MOS data taken in 2006 was also eliminated for the same reason.

The MOS event files were processed using the pipeline tool \texttt{emchain} in version 18.0.0 of the XMM-Newton Science Analysis System (\texttt{SAS}). 
The RGS data were processed with the RGS pipeline tool \texttt{rgsproc} in \texttt{SAS}. 
We obtained total  exposure times as summarized in Table~\ref{obs}  after removing background flares according to the screening with the standard event selection criteria.

 \begin{deluxetable*}{c|clcccc}
\tablenum{1}
\tablecaption{Observation data}
\label{obs}
\tablewidth{0pt}
\tablehead{
& \colhead{Observation ID} & \colhead{Starting Time} & \multicolumn4c{Effective Exposure Time (ks)} \\
& \colhead{} & \colhead{} & \colhead{MOS1} & \colhead{MOS2} & \colhead{RGS1} & \colhead{RGS2} }
\startdata  
  RCW~103&0113050601 & 2001 Sep 3 & 13.2 & 14.1 & 9.9 & 9.6 \\
  &0302390101 & 2005 Aug 23  & 54.5 & 56.5 & 58.7 & 58.6 \\
  &0743750201 & 2016 Aug 19 & \nodata & \nodata & 49.7 & 48.8 \\
  &0805050101 & 2018 Feb 14 & 55.7 & 57.4 & 35.6 & 35.1 \\
  Background&0113050701 & 2001 Sep 3 & 10.7 & 11.0 & 11.0 & 10.7 \\
\enddata
\end{deluxetable*}

\section{Analysis and Results}\label{sec:ana_res}
\begin{figure}[ht]
 \begin{center}
  \includegraphics[width=80mm]{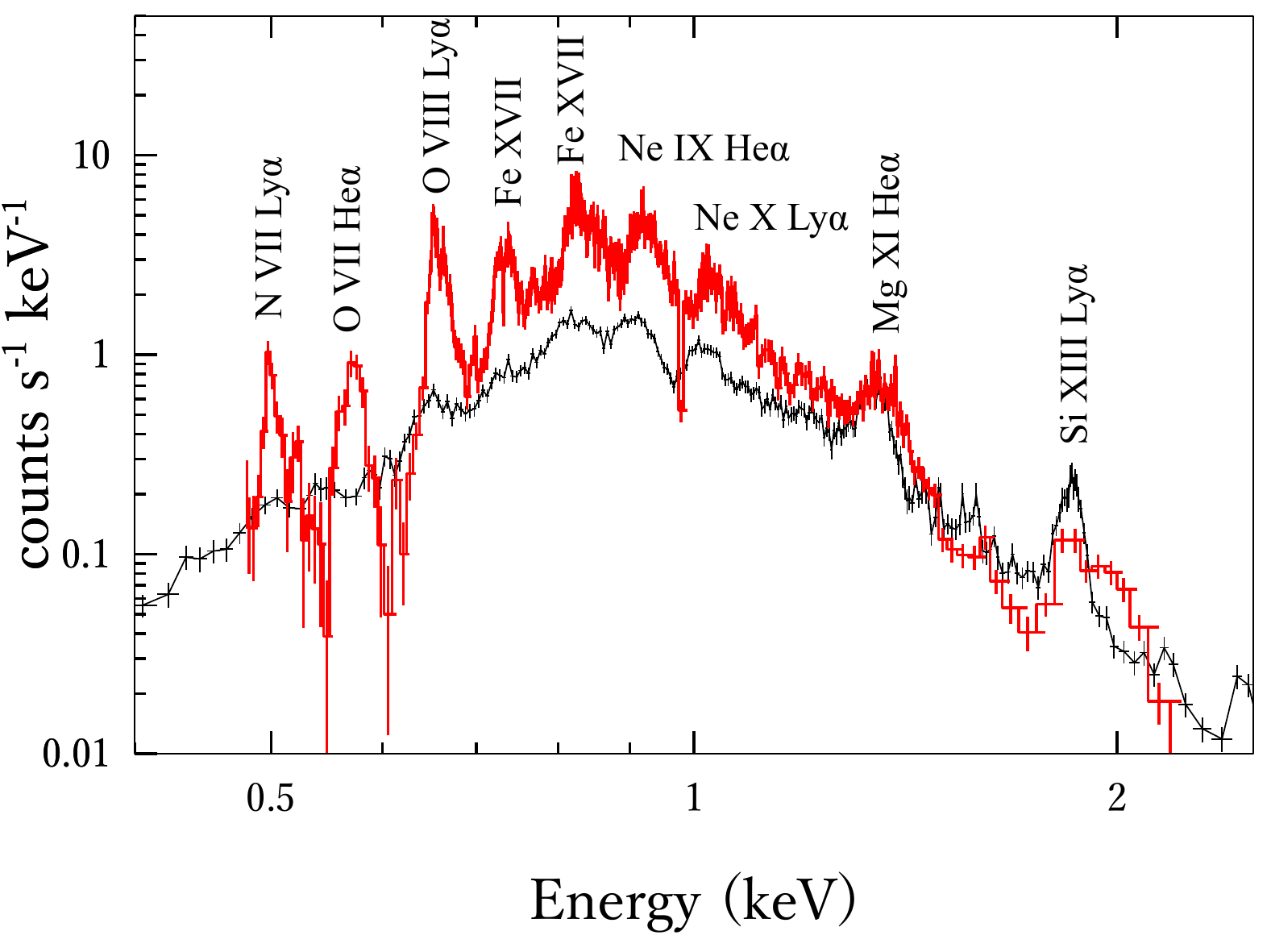}
 \end{center}
 \caption{MOS and first-order RGS spectra of region~A observed in 2001. The red lines and points represent the RGS data whereas black represents that of MOS1.}
 \label{spec}
\end{figure}

As shown in the soft-band (0.5--1.2~keV) image of RCW~103 (Figure~\ref{rgb}), the remnant has bright shells in the southeast and northwest, where CSM is thought to be swept up by forward shocks \citep[cf.][]{Frank_2015}.
We selected three compact bright knots in the shells (regions~A, B, and C; see Figure~\ref{rgb}) so that we can get high energy resolution with the RGS.
The angular size of each region is roughly $\lesssim1\arcmin$, which is enough to resolve CNO lines. 
From an obtained RGS spectrum shown in Figure~\ref{spec}, we detected several  fluorescence lines below 1~keV.
Among them, it is notable that a strong emission line of N$^{6+}$ (\ion{N}{7} Ly$\alpha$ at 0.50~keV) was clearly identified for the first time in RCW~103.

\begin{figure*}[ht]\label{rgb}
 \begin{center}
  \includegraphics[width=160mm]{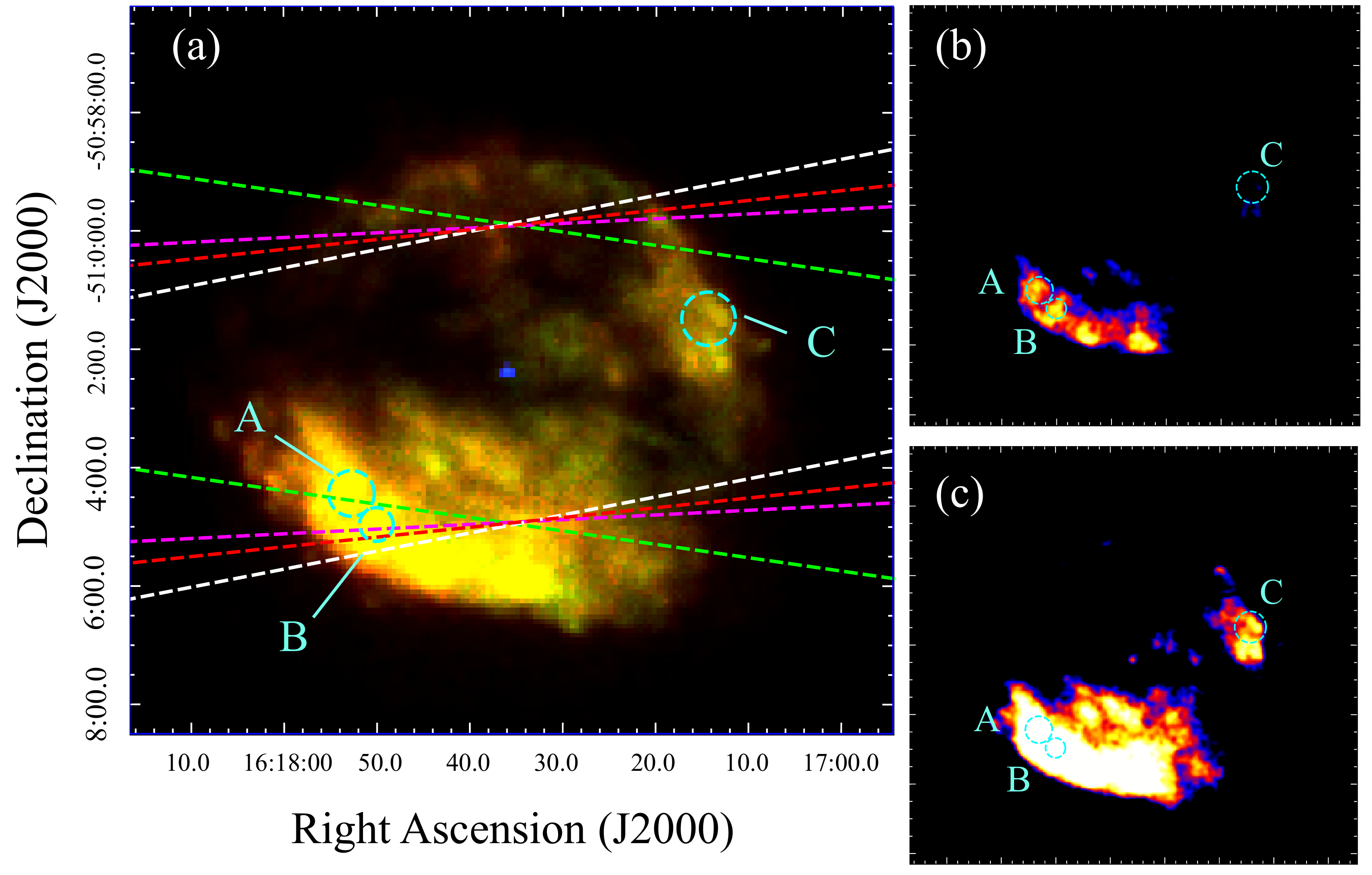}
 \end{center}
 \caption{(a): True-color image of RCW~103 from  the MOS1+MOS2 data taken in 2001. Red, green and blue  correspond to the energy band of 0.40--0.75~keV, 0.75--1.3~keV and 2.0--7.2~keV, respectively. The dashed lines represent the cross-dispersion widths of the RGS (5$\arcmin$) and each color corresponds to  observation IDs of 0113050601 (white), 0302390101 (red), 0743750201 (green), and 0805050101 (magenta). Dashed cyan circles represent the spectral extraction regions. (b): Soft-band image of RCW~103 (0.5--1.2~keV). (c): The same as panel (b) but the color scale is different for clarifying a faint structure in Region~C.}
 \label{rgb}
\end{figure*}

\begin{figure*}[ht]
 \begin{center}
  \includegraphics[width=180mm]{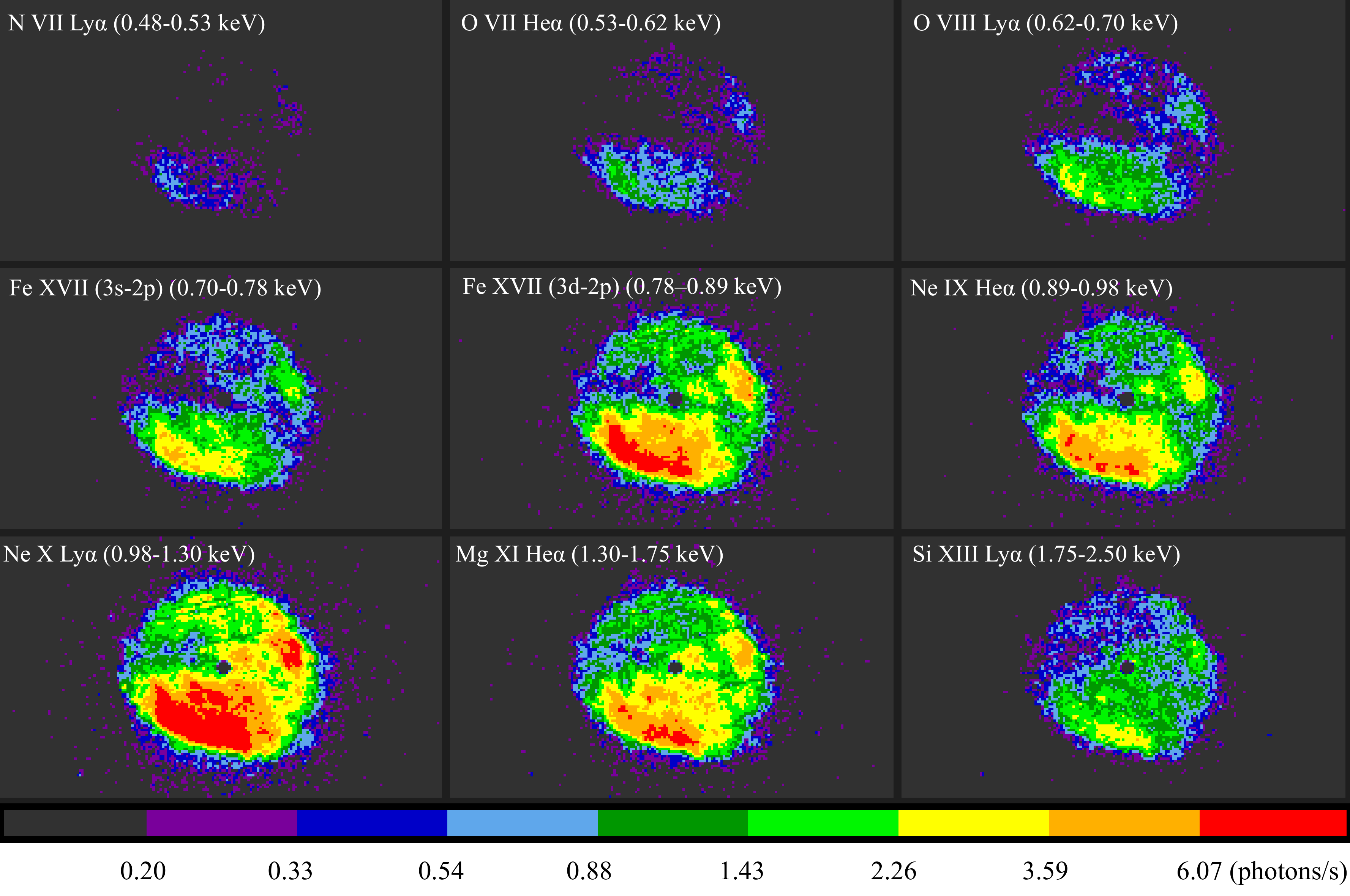}
 \end{center}
 \caption{MOS1 images showing the brightness distribution of each emission lines labeled in Figure~\ref{spec}. The center region, where  the bright magnetar is located, is masked  for displaying purpose.}
 \label{line}
\end{figure*}

For the following spectral analysis, we used version 12.11.1 of the \texttt{XSPEC} software \citep{Arnaud_1996}, in which we use the maximum likelihood W-statistic \citep{Wachter_1979}. 
We combined the RGS data (RGS1$+$2) and simultaneously fitted first/second-order RGS and MOS spectra.
Soft ($\leq$1~keV) and full ($\leq$5~keV) energy bands were used for the RGS and MOS, respectively.
We modified RGS response matrices (RMFs) so as to take into account the spatial degradation of the resolution with \texttt{rgsrmfsmooth} in \texttt{FTOOL}.
Since each line image has different spatial distributions (Figure~\ref{line}), we applied different RMFs to each line band.

Previous studies \citep{Frank_2015, Braun_2019, Zhou_2019} report that the X-ray emission of RCW~103 generally originates both from shock-heated ejecta and CSM (or  in some cases interstellar medium; ISM).
We therefore applied an absorbed two-component non-equilibrium ionization (NEI) model with variable abundance in XSPEC: \verb|tbabs*(vnei+vnei)|, where \verb|tbabs| represents the T\"{u}bingen-Boulder interstellar absorption \citep{Wilms_2000}.
The hydrogen column density ($N_{\rm{H}}$) was allowed to vary. 
Free parameters of the NEI components were the electron temperature ($kT_{\rm{e}}$), ionization timescale ($n_{\rm{e}}t$, where $n_{\rm{e}}$ is the electron number density and $t$ is the elapsed time since ionization started), and normalization. 
The abundances of Ne, Mg, Si, S, and Fe (=Ni) for the hot component and those of N, O, and Ne for the cold component were set free, whereas the others were fixed at solar values \citep{Wilms_2000}. 

We note that the above model cannot represent a line-like residuals found at $\sim1.2$~keV, which is often pointed out by previous SNR studies using the NEI model \citep[cf.][]{Okon_2020}.
They suggest that it is attributed to an uncertainty in emissivity data of Fe-L lines, or to physical processes such as charge exchange that are not taken into account the model.
According to their method, we added a Gaussian component at 1.23~keV and found that the fit was improved (W-statistics value/d.o.f.: 20852/15341 to 20769/15339 for region A) without any significant change of the other parameters.
We thus applied the two-component NEI model with a single Gaussian for the following analysis.

\begin{figure*}[ht]
 \begin{center}
  \includegraphics[width=180mm]{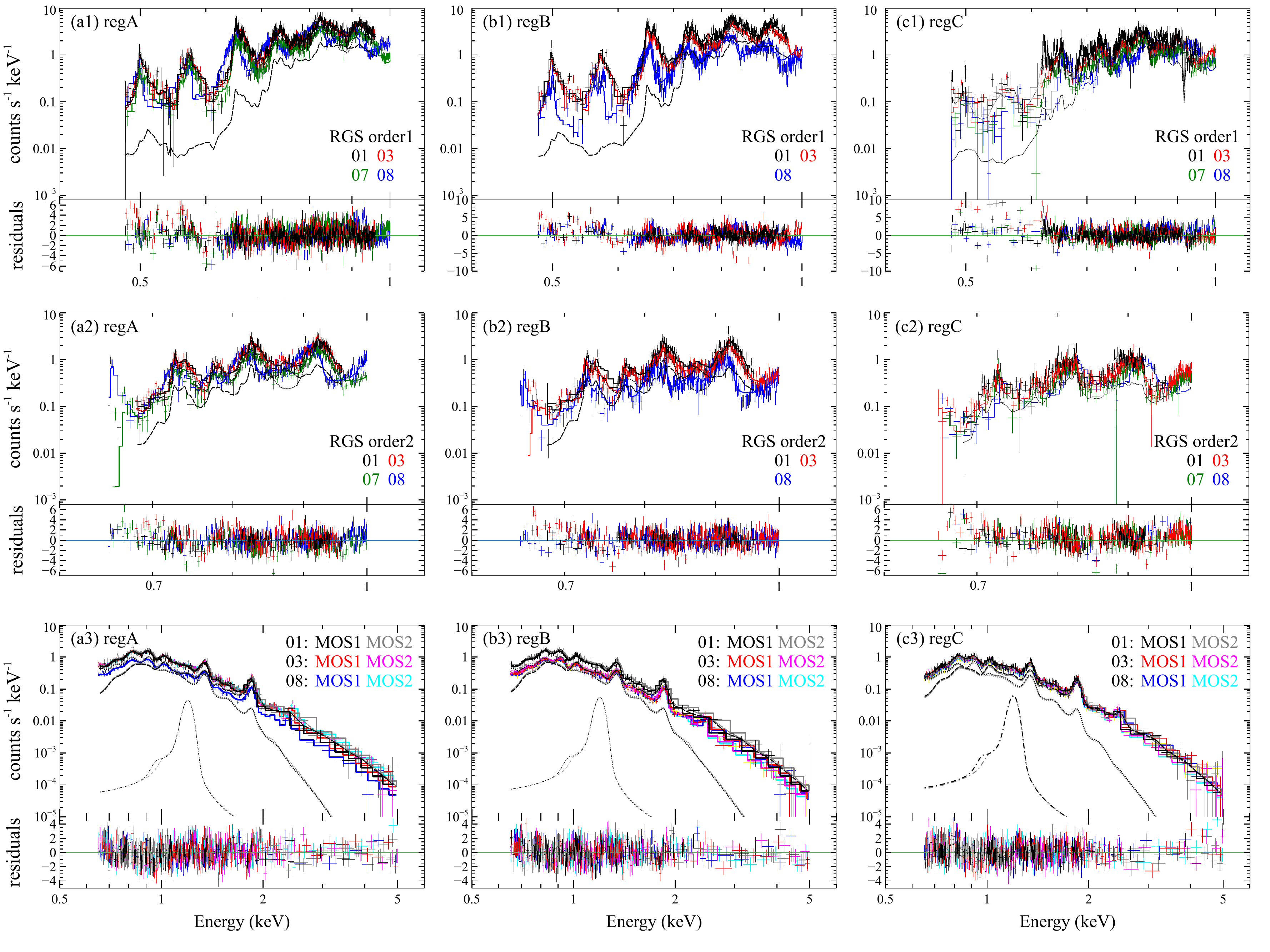}
 \end{center}
 \caption{X-ray Spectra of all the selected regions. \textit{Left}: Spectra of region~A obtained with the first- (a1) and second- (a2) order of RGS1+2 and those of MOS1/2 (a3). The numbers 01, 03, 07 and 08 represent the observation IDs 0113050601, 0302390101, 0743750201, and 0805050101, respectively. The dashed  and dotted lines are the best-fit NEI components with high and low $kT_{\rm{e}}$, respectively. The dot-dashed lines show the Gaussian component (see text). \textit{Middle and Right}: The same as the left but  spectra of regions~B and C.}
 \label{speca}
\end{figure*}

\begin{deluxetable*}{llccc}
\tablenum{2}
\tablecaption{Best-fit parameters of the spectrum}
\label{par}
\tablewidth{0pt}
\tablehead{
\colhead{Component} & \colhead{Parametars (unit)} & \colhead{regA} & \colhead{regB} & \colhead{regC}}
\startdata  
Absorption & $N_{\rm{H}}$ (10$^{22}$~cm$^{-2}$)& $0.881^{+0.002}_{-0.004}$ & $0.91\pm{0.01}$  &  $1.07\pm{0.01}$ \\
Low-temperature NEI & $kT_{\rm{e}}$ (keV) & $0.202^{+0.003}_{-0.001}$  & $0.200\pm{0.001}$ & $0.198\pm{0.002}$ \\
(CSM)& N & $1.4\pm{0.1}$ & $1.5\pm{0.1}$ & $1.1^{+0.2}_{-0.3}$ \\
& O & $0.38\pm{0.01}$ & $0.42^{+0.03}_{-0.01}$ & $0.33^{+0.02}_{-0.03}$ \\
& Ne & $0.62^{+0.01}_{-0.02}$ & $0.65\pm{0.02}$ & $0.57^{+0.05}_{-0.01}$  \\
& $n_{\rm{e}}t~(10^{11}\ \rm{cm^{-3}~s})$ & $>100$ & $>100$ & $>100$   \\
& norm & $1.23^{+0.05}_{-0.04}$  & $1.31^{+0.07}_{-0.13}$ & $0.51^{+0.06}_{-0.03}$ \\
High-temperature NEI & $kT_{\rm{e}}$ (keV) & $0.650^{+0.003}_{-0.007}$ & $0.59\pm{0.01}$ & $0.57\pm{0.01}$ \\
(ejecta)& Ne & $2.0\pm{0.1}$ & $1.8^{+0.6}_{-0.2}$ & $1.2\pm{0.1}$ \\
& Mg & $1.5\pm{0.1}$ & $1.5^{+0.4}_{-0.1}$ & $1.4^{+0.2}_{-0.1}$ \\
& Si & $2.2^{+0.1}_{-0.2}$ & $2.0\pm{0.2}$ & $1.9^{+0.2}_{-0.1}$ \\
& S & $1.3^{+0.2}_{-0.1}$ & $1.4\pm{0.2}$ & $1.3^{+0.2}_{-0.1}$ \\
& Fe (=Ni) & $1.48^{+0.2}_{-0.3}$ & $1.6^{+0.2}_{-0.1}$ & $1.3\pm{0.1}$ \\
& $n_{\rm{e}}t~(10^{11}\ \rm{cm^{-3}~s})$ & $3.2^{+0.4}_{-0.2}$ &  $5.0^{+1.7}_{-0.3}$ & $3.7^{+0.3}_{-0.7}$ \\
& norm & $0.044^{+0.003}_{-0.002}$ & $0.055\pm{0.005}$ & $0.023\pm{0.002}$ \\  
\hline
& W-statistic/d.o.f. & 20769/15339 & 13992/12746 & 15517/15340 \\
\hline
\enddata
\end{deluxetable*}

Figure~\ref{speca} shows fits of the two-component NEI model to the spectra.
The best-fit parameters are listed in Table~\ref{par}.
The values of $N_{\rm{H}}$ and parameters for the high-$kT_{\rm{e}}$ component  are consistent with the ejecta abundance measured by previous studies \citep{Frank_2015, Braun_2019, Zhou_2019}.
We find that there is no significant difference among the three regions in the abundances of the low-$kT_{\rm{e}}$ component.
The N abundance particularly exceeds the solar value in at least two regions.
On the other hand, those of O and Ne are  significantly lower than solar.
These results suggest that the detected N and O lines are originated from the shock-heated CSM blown off by the progenitor’s stellar wind.
 
\section{Discussion}
\subsection{Evidence of CSM and its composition}\label{sec:evi_com}
While it is expected that CSM contains N-rich materials produced in stars \citep[e.g.,][]{Maeder_1983} and that the resultant N/O becomes higher than solar, it has been difficult to detect N lines in SNRs so far.
\cite{Frank_2015} found that the plasmas in the shell of RCW~103 have sub-solar ($\sim0.5$) abundances of heavier elements (Ne, Mg, Si, S, and Fe) and presumed that the CSM is a largely dominant component across the remnant.
Our result confirms their conclusion and shows robust evidence of the N-rich shell in RCW~103 for the first time.
The averaged abundance ratio is $\rm{N/O}=3.8\pm0.1$, which fits well into the picture that the N-rich materials yielded from a massive star are blown out at the RSG or WR stage \citep[e.g.,][]{Owocki_2004}.

\begin{figure*}[ht]
 \begin{center}
  \includegraphics[width=160mm]{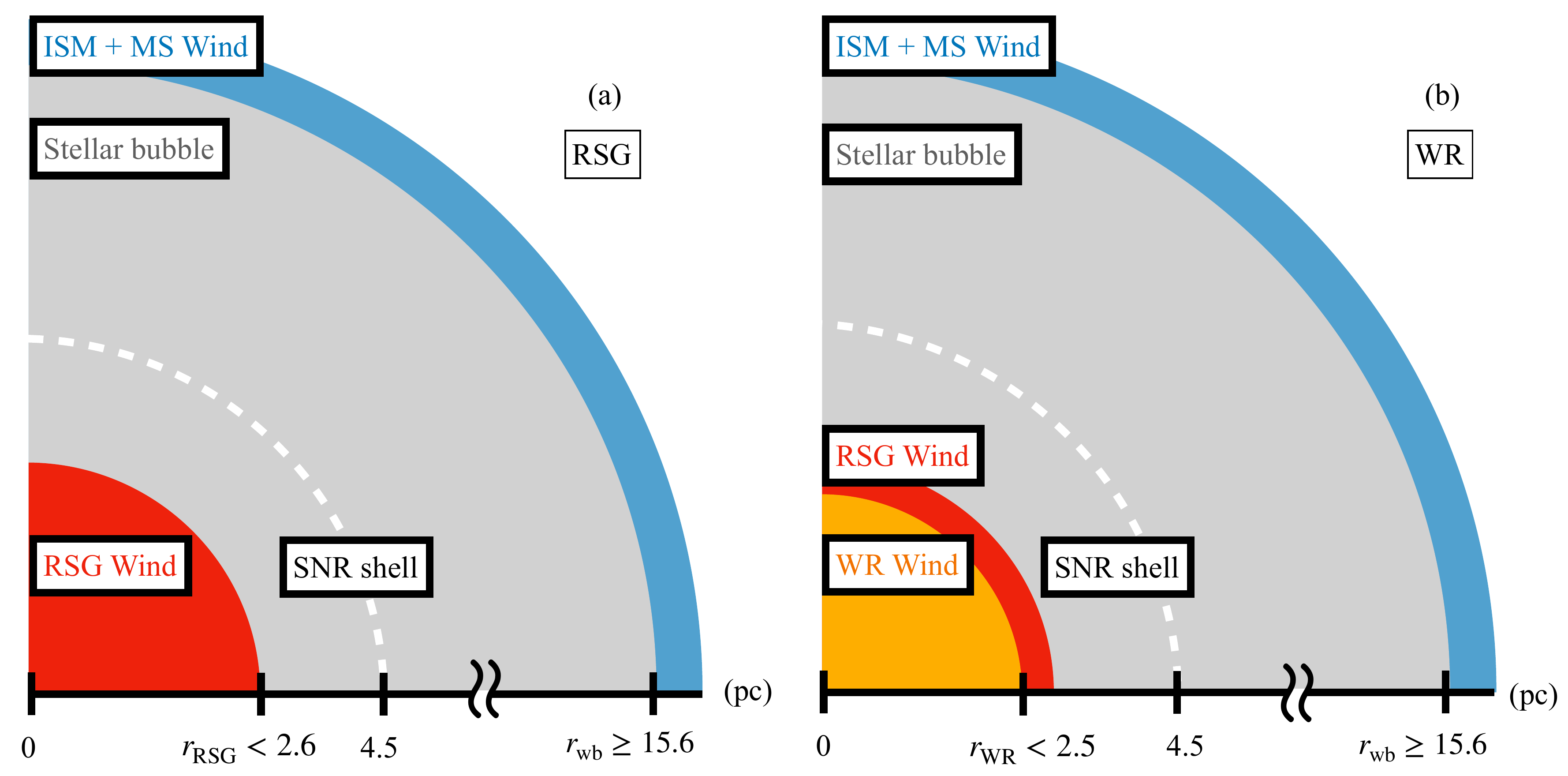}
 \end{center}
 \caption{Schematic view of an environment around a star at the end of its life. \textit{Left (a)}: Case of a low-mass star ending up as a RSG. Colors correspond to  the shell of ISM and MS winds (blue), stellar bubble (grey), and RSG winds (red). White dashed line represents the position of the forward shock of RCW~103. \textit{Right (b)}: Case of a massive star ending its life as a WR. The region dominated by a WR wind is represented as orange.}
\label{dis}
\end{figure*}

On the basis of the result, we constrain environments around the progenitor of RCW~103 in its pre-supernova phase, where stellar winds are blown out with various abundances in various conditions.
As illustrated in Figure~\ref{dis}, we assume two cases with different CSM structures, in which we take into account CSM-hydrodynamic and stellar evolutions as described below.
An outermost region from a central star is called a main sequence (MS) shell, which contains MS winds and swept-up ISM \citep{Weaver_1977}.
The region inside the MS shell is called a stellar wind bubble formed by shock waves through collision between the MS winds and the ISM \citep{Castor_1975}.

When the wind bubble comes into pressure equilibrium with surroundings, its radius $r_{\rm{wb}}$ is described as
\begin{equation}\label{eq_rb}
\begin{split}
r_{\rm{wb}} =~&5.8\left(\frac{M_{\rm{w}}}{10^{-2}~M_{\odot}}\right)^{1/3}\left(\frac{v_{\rm{w}}}{700~\rm{km~s^{-1}}}\right)^{2/3} \\
&\times\left(\frac{p_{\rm{i}}/k}{10^4~\rm{cm^{-3}~K}}\right)^{-1/3}~\rm{pc},
\end{split}
\end{equation}
where $M_{\rm{w}}$ and $v_{\rm{w}}$ are total mass and velocity of the MS wind, respectively.
The value $p_{\rm{i}}$ is the pressure of ISM, and $k$ is Boltzmann's constant \citep{Chevalier_1999}. 
In our Galaxy, $p_{\rm{i}}/k$ and $v_{\rm{w}}$ at MS are typically $\sim10^4~\rm{cm^{-3}~K}$ \citep[e.g.,][]{Bergh_1998} and $\sim10^3~\rm{km~s^{-1}}$, respectively \citep[as summarized by][]{Chen_2013}.
Since the total wind mass $M_{\rm{w}}$ depends on a progenitor mass \m, we calculated a typical value using a stellar evolution code (HOngo Stellar Hydrodynamics Investigator; HOSHI) developed by \cite{Takahashi_2013, Takahashi_2014, Takahashi_2016, Takahashi_2018, Yoshida_2019}.  
As a result, if the progenitor has \m \ of $10M_{\odot}$, obtained wind mass is $M_{\rm{w}}~=~9.5\times10^{-1}~M_{\odot}$.
In this condition, the wind bubble radius $r_{\rm{wb}}$ is calculated as 
\begin{equation}\label{eq_rb_zams}
\begin{split}
\left.r_{\rm{wb}}\right|_{M_{\rm ZAMS}=10M_{\odot}} =~&15.6\left(\frac{M_{\rm{w}}}{9.5\times10^{-1}~M_{\odot}}\right)^{1/3}\\
\times\left(\frac{v_{\rm{w}}}{10^{3}~\rm{km~s^{-1}}}\right)^{2/3}&\left(\frac{p_{\rm{i}}/k}{10^4~\rm{cm^{-3}~K}}\right)^{-1/3}~\rm{pc}.
\end{split}
\end{equation}
Since the size of the wind bubble has a positive liner relation with \m \ \citep{Chen_2013}, Eq.\ref{eq_rb_zams} gives a minimum value of $r_{\rm{wb}}$ by assuming a lower limit $M_{\rm{ZAMS}}\sim10M_{\odot}$ for CC-SN progenitors.
On the other hand, RCW~103 has a radius of $4.5~\rm{pc}$ (see Figure~\ref{dis}), which confirms that the bright clumps observed in our study are within the MS wind shell; the low-$kT_{\rm{e}}$ component is not contaminated by ISM.

To investigate an abundance pattern of the swept-up CSM, a stellar wind blown out  just before the SN explosion is the most important component.
Relatively low \m \ stars end up as RSG, surrounded by an RSG wind (panel (a) of Figure~\ref{dis}), whose radius $r_{\rm{RSG}}$ is described as
 \begin{equation}\label{eq_rR}
\begin{split}
r_{\rm{RSG}} \leq~&2.6\left(\frac{{\dot{M}}_{\rm{w}}}{5\times10^{-7}~M_{\odot}~\rm{yr^{-1}}}\right)^{1/2}\left(\frac{v_{\rm{w}}}{15~\rm{km~s^{-1}}}\right)^{1/2} \\
&\times\left(\frac{p_{\rm{wb}}/k}{10^4~\rm{cm^{-3}~K}}\right)^{-1/2}~\rm{pc},
\end{split}
\end{equation}
where ${\dot{M}}_{\rm{w}}$ and $v_{\rm{w}}$ are the mass-loss rate and velocity of the RSG wind, respectively.
The value $p_{\rm{wb}}$ represents the pressure of stellar wind bubbles \citep{Chevalier_Emmering_1989, Chevalier_2005}. 
In Eq.\ref{eq_rR}, we assume a typical pressure of bubbles $\sim10^4~\rm{cm^{-3}~K}$ \citep{Castor_1975}, wind velocity at the RSG phase $10\textrm{--}20~\rm{km~s^{-1}}$ \citep{Goldman_2017}, and  the mass-loss rate $\sim10^{-5\textrm{--}7}~M_{\odot}~\rm{yr^{-1}}$ calculated with the HOSHI code.
The result indicates that the swept-up CSM in RCW~103 contains all RSG-wind materials if the progenitor ended up as RSG.

In case that the progenitor of RCW~103 was a high-\m \ star, we additionally need to consider an effect of the WR wind, which may sweep up the ambient RSG wind material before the SN explosion.
The WR wind expands into the RSG-wind materials with a velocity of 100--200~km~s$^{-1}$ \citep{Chevalier_Imamura_1983, Chevalier_2005} during a WR phase duration of $\sim10^{3\textrm{--}4}$ year for \m~$\leq60~M_{\odot}$ \citep[from a calculation using a stellar evolution model Geneva Code][]{Ekstrom_2012}.
From these estimations, the radius of the shell of the RSG wind  swept up by the WR wind is calculated to be $\leq2.5~\rm{pc}$.
The bright clumps therefore contain all the RSG and WR materials if the progenitor of RCW~103 ended up as a WR star (panel (b) of Figure~\ref{dis}).

\subsection{Stellar Wind History of the Progenitor}\label{sec:his}
\begin{figure*}[ht]
 \begin{center}
  \includegraphics[width=180mm]{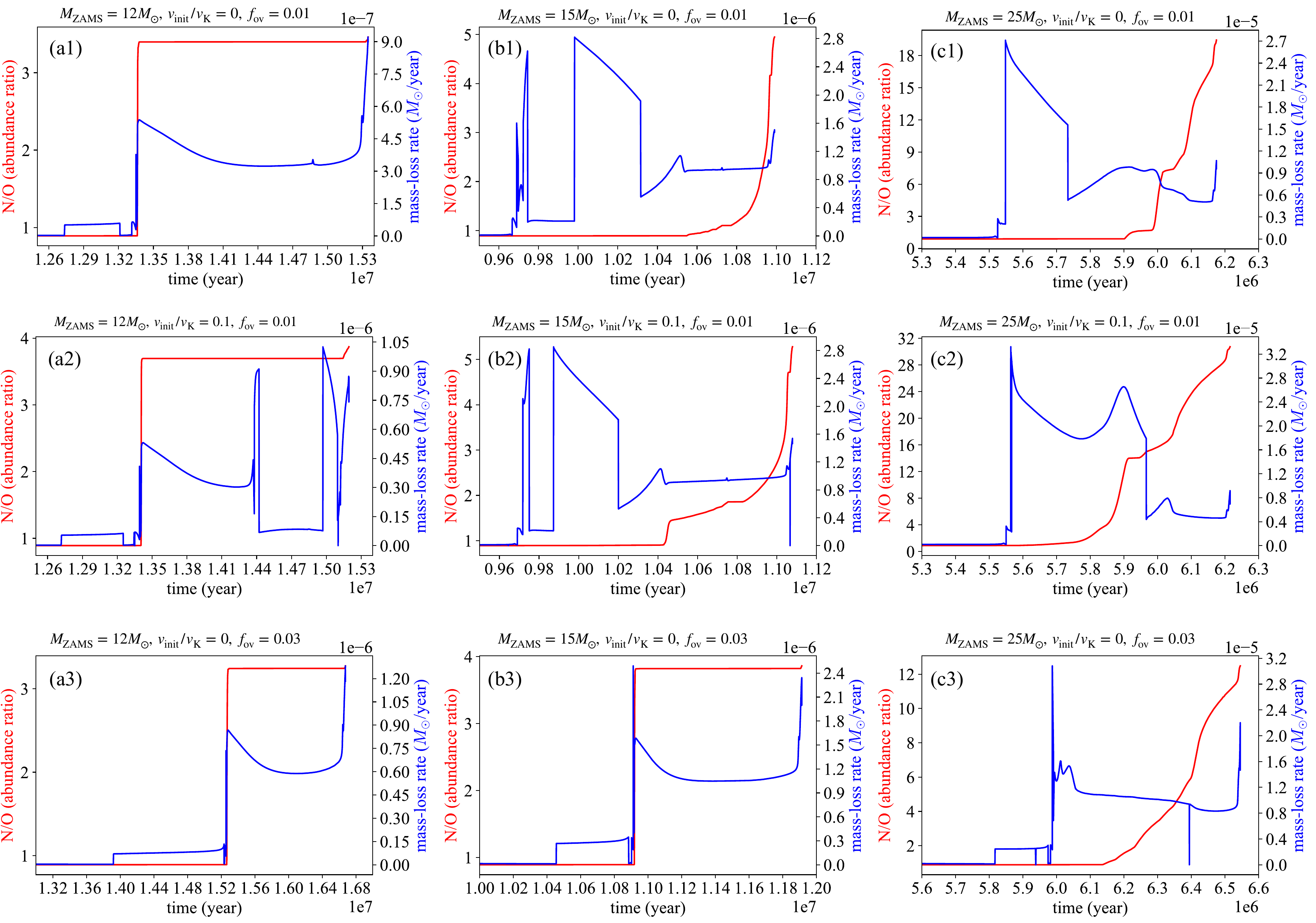}
 \end{center}
 \caption{Relation between the time from the birth of stars and the abundance ratio of N/O in the stellar surface (red) and the mass-loss rate (blue). \textit{Top}: models with 12$M_{\odot}$ (a1), 15$M_{\odot}$ (b1), and 25$M_{\odot}$  (c1), with no rotation and a low overshoot parameter ($f_{\rm{ov}}$ = 0.01). \textit{Middle}: Same as the top but the initial rotation velocity is higher ($v_{\rm{init}}/v_{\rm{K}}=0.1$). \textit{Bottom}: Same as the top but the overshoot parameter is higher ($f_{\rm{ov}}$ = 0.03).}
\label{history}
\end{figure*}

For constraining progenitors, we need to know a relationship between  CSM abundances and conditions of stars. 
The N/O ratio measured in CSM is related to the progenitor \m, rotation velocity \citep{Maeder_2014}, and convective overshoot (the extrusion of convective motion due to non-zero velocities of material on a boundary between the convection zone and the radiative zone in stars). 
 We, therefore, investigate trends between N/O and progenitor parameters by using the HOSHI code.
 The results are summarized in Figure~\ref{history}.
 Note that in our calculation, the metallicity, which may also affects our study \citep{Maeder_2014}, is assumed to be a solar since RCW~103 is one of young Galactic SNRs.
 As shown in Figure~\ref{history}, N/O increases in the stellar surface during the evolution, but its timescale highly depends on the assumed stellar properties. 
 
 \m \ is generally  thought to have a positive relationship with N/O since the mass-loss rate is positively related to \m~\citep[e.g.,][]{Mauron_2011}.
 Figures~\ref{history} (b1) and (c1) indicate that a more rapid increase of N/O is expected in case of higher \m.
 This is because a sub-solar outer layer ($\rm{N/O}\sim1$) is blown out in early stage of evolution due to the higher mass loss rate caused by stronger radiative force \citep[e.g.,][]{Mauron_2011}. 
 In contrast, lower mass progenitors (\m$\leq12M_{\odot}$) result in a rapid increase of N/O at an earlier stage of stellar evolution.
 This is possibly because N-rich materials in the H-burning layer of stars with \m$=10\textrm{--}12~M_{\odot}$ are carried to the stellar surface by convection at earlier evolution stage than those of more massive stars: for instance, in case of  \m$\sim15~M_{\odot}$, the He burning occurs before the RSG phase, which prevents the star's expansion resulting in an inefficient transfer of N to the stellar surface.

As claimed by previous studies \citep{Owocki_2004, Heger_2000}, rotation velocities of stars also affect the final yield of the CNO elements.
This is, for instance, clearly seen in Figures~\ref{history} (b1) and (b2), where N/O increases more rapidly by a faster rotation ($v_{\rm{init}}/v_{\rm{K}}=0.1$, where $v_{\rm{init}}$ is the initial surface rotation velocity and $v_{\rm{K}}$ is the Kepler velocity at the surface). 
The trend is interpreted as an effect of the centrifugal force, which enhances the mass-loss rate at the stage of OB stars \citep{Owocki_2004} and carries materials from the He-burning layer to the H-burning layer such as C and O causing an additional N production by CNO-cycle \citep{Heger_2000}.

We also investigate an effect of the convective overshoots, which may also enhance the abundance ratio of N/O because convective layers contain large amount of N-rich materials \citep[e.g.,][]{Luo_2022}.
In the HOSHI code,  convective overshoot is treated as a diffusion approximation with the diffusion coefficient $D = D_0 \exp(-2{\Delta}r/f_{\rm{ov}}H_{\rm{p}}$), where $D_0$ is the diffusion constant at the boundary, ${\Delta}r$ is the distance from the boundary, $f_{\rm{ov}}$ is the overshoot parameter variable, and $H_{\rm{p}}$ is the pressure scale hight. 
We tried two overshoot models with different variables \citep[$f_{\rm{ov}}=0.01$ and 0.03, namely model $M_A$ and $L_A$, respectively, given in \S~2 of][]{Luo_2022} and found that a higher overshoot parameter results in a higher N/O on the stellar surface as seen in panels (b1)  and (b3) of Figures~\ref{history}.
This is because N-rich materials in the H-burning layer are carried to the region closer to the surface by convection. 

\subsection{Progenitor constraint with N/O}\label{sec:pro}
\begin{figure}[ht]
 \begin{center}
  \includegraphics[width=85mm]{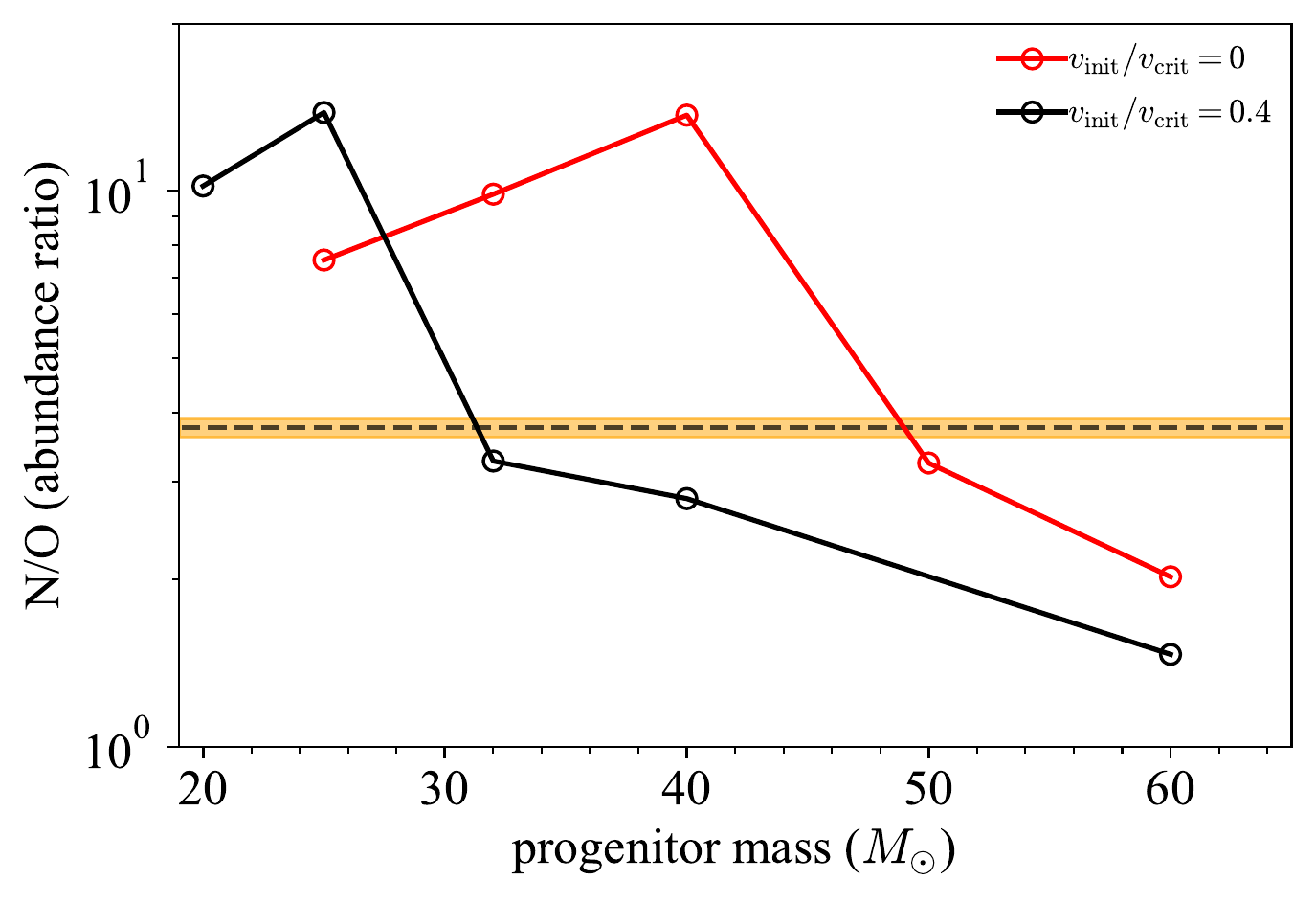}
 \end{center}
 \caption{Measured N/O from our observation  (the orange-hatched region)  and those expected  in the swept-up CSM with different \m \ calculated with Geneva code (the red and black circles). The black dashed horizontal line and the width show the averaged N/O in RCW~103 and its $1\sigma$ errors, respectively.
The red and black circles correspond to cases of $v_{init}/v_{crit}$ = 0 and 0.4, respectively.}
\label{NO_wr}
\end{figure}

\begin{figure}[ht]
 \begin{center}
  \includegraphics[width=85mm]{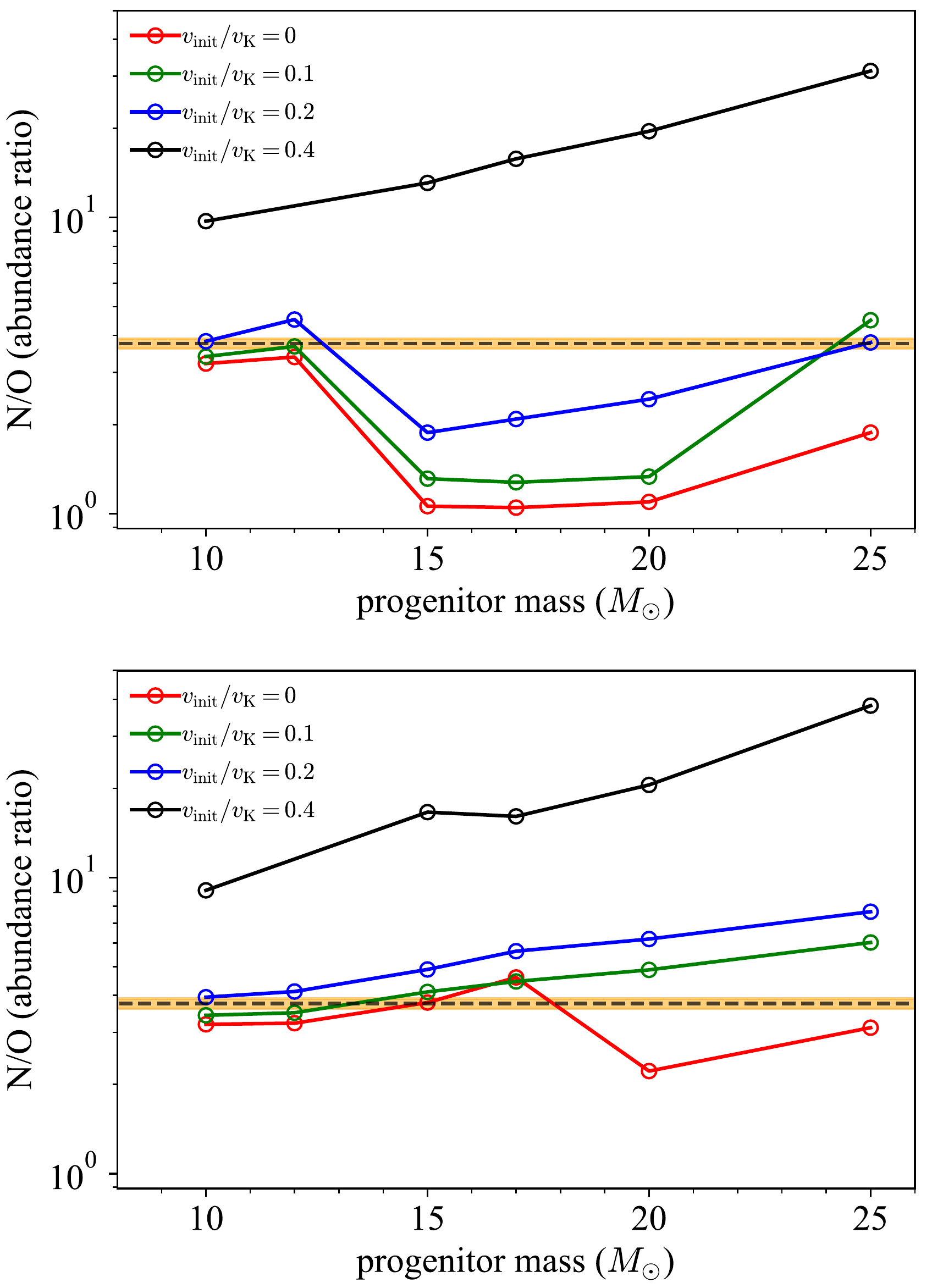}
 \end{center}
 \caption{Same as Figure~\ref{NO_wr} but for the RSG models calculated with the HOSHI code. Colors correspond to different rotation velocities; $v_{init}/v_{K}$ = 0 (red), 0.1 (green), 0.2 (blue), 0.4 (black). Models with lower ($f_{\rm{ov}}$ = 0.01) and higher ($f_{\rm{ov}}$ = 0.03) overshoot parameters are displayed in the top and bottom panels, respectively.}
\label{NO_rsg}
\end{figure}

In accordance with the calculations demonstrated in \S~\ref{sec:his}, we compared the measured abundance ratio of N/O$=3.8\pm0.1$ in the CSM of RCW~103 with several models having different \m, rotation velocity, and overshoot parameters.
In this analysis, we calculated the abundance ratio of N/O by summing up a total amount of N and O contained in post-main sequence winds.
The abundance evolution of WR stars and RSGs was obtained from the results in \cite{Ekstrom_2012} and the model calculations using Hoshi code, respectively.
The results are displayed in Figure~\ref{NO_wr} (WR) and Figure~\ref{NO_rsg} (RSG), from which we can constrain the progenitor properties and origin of RCW~103.

If we assume the WR models, the progenitor of RCW~103 is likely to be a 25--32$M_{\odot}$ star with a very rapid initial rotation velocity ($v_{\rm{init}}/v_{\rm{crit}}=0.4$, where $v_{\rm{crit}}$ is the critical velocity reached when the gravitational acceleration is equal to the centrifugal force in the Roche model) or a 40--50$M_{\odot}$ star with no rotation.
Here, it is noted that $v_{\rm{crit}}$ is slightly different to $v_{\rm{K}}$ by a factor of $\sqrt{2/3}$.
These results are, however, inconsistent with recent theoretical expectations that stars over $20M_{\odot}$ are hard to explode and that those over $35M_{\odot}$ mostly become black holes  \citep{Sukhbold_2016}.
Although the explodability of such massive stars is under debate, we conclude that it is less likely  that the progenitor of RCW~103 was a WR star.
 
 In the case of RSG (Figure~\ref{NO_rsg}), the progenitor of RCW~103 is likely to be a low-mass ($\leq12M_{\odot}$) or medium-mass $\geq15M_{\odot}$) star with normal rotation velocities ($v_{init}/v_{K}\leq0.2$).
 The latter result  is consistent with a previous estimation based on the ejecta abundances \citep[$\sim18M_{\odot}$;][]{Frank_2015}.
It is, however, somewhat unlikely in the context of birth conditions of compact objects \citep{Higgins_Vink_2019}.
They present a relation between the progenitor properties and a compactness parameter $\zeta_M$, which is defined  as
\begin{equation}\label{eq_com}
\begin{split}
\zeta_M =\left.\frac{M/M_{\odot}}{R(M_{\rm{bary}}=M)/1000~\rm{km}}\right|_{t=t_{\rm{bounce}}},
\end{split}
\end{equation}
where $M$ is the relevant mass scale for black hole formation and $R(M_{\rm{bary}}=M)$ is the radial coordinate that encloses $M$ at the time of core bounce \citep{Oconnor_Ott_2011}.
\cite{Higgins_Vink_2019} claimed that $\zeta_M$  is related to how easily a pre-supernova stellar core explodes.
As they suggest, it is difficult for medium \m \ (15--40$M_{\odot}$) stars to explode except stars having very rapid rotation velocities ($v_{init}/v_{K}\geq0.4$), which can be ruled out by our estimation 
(the black lines in Figure~\ref{NO_rsg}). 

By contrast, low-mass (10--12$M_{\odot}$) progenitor models are acceptable  from the point of view of the explodability and also well fit with previous estimations derived from a comparison of metal compositions of ejecta with a core-collapse nucleosynthesis model; 12--13~$M_{\odot}$ \citep{Braun_2019} and  $<$~13~$M_{\odot}$ \citep{Zhou_2019}. 
We also point out that according to a relation between Fe/Si and \m \  given by \citet{Katsuda_2018} the obtained ejecta abundance ratio ($\rm{Fe/Si}>0.6$ from Table~\ref{par}) prefers a lower-mass star as well.
From these results, we  conclude that the progenitor of RCW~103  was  most likely a  star with \m \ of 10--12$M_{\odot}$.
We also presume that its rotation velocity was relatively normal; $v_{init}/v_{K}\leq0.2$.

\subsection{Origin of the Magnetar and Future Perspectives}\label{sec:mag}
 RCW~103 has a magnetar candidate 1E~161348$-$5055.
Our result may also hint at  physical conditions that lead to form magnetars after SN explosions.
One of the well-established hypotheses is a ''dynamo effect'' model \citep{Thompson_Duncan_1993}.
In this scenario,  magnetars are born with rapidly rotating (on the order of millisecond) proto-neutron stars (PNSs), which can power energetic SNe (or release most of the energy through gravitational waves).\par
Since the rotation velocity of PNSs shows a positive correlation with  \m, \cite{Gaensler_2005} and \cite{Heger_2005} indicate that fairly high \m \ ($\geq~35~M_{\odot}$) is required to produce a magnetar with this effect in the case of the progenitors with $v_{init} =200~\rm{km~s^{-1}}$.
On the other hand, \cite{White_2022} suggests that low mass progenitors (\m~=9--25~$M_{\odot}$) with no initial rotation can also make the strong magnetic fields by the dynamo effect considering sufficiently low Rossby number.
\cite{Masada_2022} also suggests that the strong magnetic fields can be generated in the relatively slowly ($\sim 100~\rm{ms}$) rotating PNSs considering different internal structures of the convection.
From Figure~\ref{NO_wr}, there is no possible solution for $\geq$~35~$M_{\odot}$ with a rapid rotation ($v_{init} > 300~\rm{km~s^{-1}}$) and the progenitor parameters given by \cite{White_2022} are consistent with our estimation.
 An alternative scenario to account for  the formation of the magnetars is the fossil field hypothesis, which requires a progenitor star with strong magnetic fields \citep{Ferrario_2006, Vink_2006, Hu_2009} originating from massive \citep[$>20M_{\odot}$;][]{Ferrario_2006, Ferrario_2008} or less massive progenitors \citep{Hu_2009}. 
 This case is also in good agreement with our estimation and it is consistent with the previous study \citep{Zhou_2019}.

Note that 1E~161348$-$5055 has an extremely long periodicity \citep[$6.67$h;][]{DeLuca_2006} unlike most magnetars \citep[$\sim1\textrm{--}10$s;][]{Olausen_2014}.
Some magnetars with short periodicity are thought to have low-mass progenitors \citep{Zhou_2019} and many OB stars in our Galaxy have the same velocities as the progenitor of RCW~103 \citep{Simone_2007}. 
Therefore, the progenitor of RCW~103 is thought to have similar physical characteristics to majority of OB stars in our Galaxy except an extremely long periodicity of 1E~161348$-$5055. 

A measurement of N/O in CSM established in our study  is a good method to constrain progenitors of SNRs; e.g., stellar properties such as \m, initial rotation velocities, and convective overshoots.
The microcalorimeter \textit{Resolve} onboard the XRISM satellite \citep{Tahiro_2018} will be able to detect the C, N and O lines in many diffuse sources in our Galaxy.
Athena \citep{Barret_2018} and Lynx \citep{Gaskin_2019} will be able to expand the targets to Large and Small Magellanic Clouds SNRs.
With these future observatories, to measure the CNO abundances will become a key method to probe the relationship between the stellar evolution and the final fate of stars, such as supernovae and compact objects, and formation probability of magnetars, neutron stars, and black holes.

\section{Conclusions}\label{sec:con}
We performed high-resolution spectroscopy of RCW~103 with the RGS onboard XMM-Newton and detected \ion{N}{7} Ly$\alpha$ (0.50~keV) for the first time.
All the spectra are well reproduced by a two-component NEI model with different temperatures.
The obtained value of $\rm{N/O}=3.8\pm{0.1}$ indicates the presence of the shock-heated CSM and thus we investigated RCW~103's progenitor parameters such as \m, rotation velocity, and overshoot parameters by using stellar evolution codes HOSHI and Geneve.
As a result, we successfully constrained the progenitor of RCW~103 as a low-mass star (\m$=10$--$12M_{\odot}$) with a relatively normal rotation velocity ($v_{init}/v_{K}\leq0.2$).
Consequently, the dynamo effect and the fossil field are plausible for the origin of the associated magnetar 1E~161348$-$5055, and we conclude that  the progenitor of RCW~103 is similar with majority of OB stars while an extremely long periodicity of 1E~161348$-$5055 is still open question.
Our method will become a useful tool for constraining progenitor properties of Galactic and extra-Galactic SNRs with future microcalorimeter missions such as XRISM, Athena, and Lynx.

We thank Dr. Hideyuki Umeda and Dr. Jacco Vink for meaningful discussions on the stellar evolution and the high resolution spectroscopy and the anonymous referee for constructive comments.
This work is supported by JSPS/MEXT Scientific Research grant Nos. JP19K03915, JP22H01265 (H.U.), JP19H01936 (T.T.), and JP21H04493 (T.G.T and T.T.).

\bibliography{sample631}{}
\bibliographystyle{aasjournal}

\end{document}